\documentclass{iau_FM}
\usepackage{graphicx}

\title[The Open Universe and Data-driven Astronomy] %% give here short title %%
{The Open Universe and Data-driven Astronomy}

\author[Priya Shah]   %% give here short author list %%
{Priya Shah 
}

\affiliation{Department of Physics, Maulana Azad National Urdu University, 
Gachibowli,\\ Hyderabad 500 032, India \\ email: {\tt priya.hasan@gmail.com} }

\pubyear{2018}
\jname{IAU’s role on global astronomy outreach, the latest challenges and bridging different communities
} 
\editors{Sze-leung Cheung, ed.}
\begin{document}

\maketitle

\begin{abstract}
Activities related to access to astronomical facilities and data could offer an effective, entry-level path for outreach and astronomy education.  The Government of Italy proposed the Open Universe  initiative that was adopted by the United Nations Office of Outer Space Affairs. Education in astronomy  is a key method to promoting rational thinking and the scientific method \cite[Giommi \etal\ (2018)]{Giommi_etal18}. We shall discuss how new methods using available data need to be used for outreach and education to help vizualise and understand actual data. We shall show, using GAIA DR2 data, how present data analysis and visualization tools can be used to identify star clusters, moving groups and runaway stars. Thus, with this example,  real data can be used to understand stellar dynamics in the galaxy. 

\end{abstract}

\firstsection % if your document starts with a section,
\section{Introduction}
Space technologies have an impact on almost all aspects of development. Space data is expensive, funded by public money and is full of undiscovered scientific treasures.  
The United Nations Office for Outer Space Affairs (UNOOSA) works to promote international cooperation in the peaceful use and exploration of space, and in the utilisation of space science and technology for sustainable economic and social development.  \cite[Giommi \etal\ (2018)]{Giommi_etal18}  describes Open Universe which is an initiative under the auspices of COPUOS with the objective of stimulating a dramatic increase in the availability and usability of space science data, extending the potential of scientific discovery to new participants in all parts of the world and empowering global educational services\footnote{http://www.openuniverse.asi.it/}.

\section{International Virtual Observatory (IVOA)}
The Virtual Observatory (VO) is the vision that `astronomical datasets and other resources should work as a seamless whole'. Many projects and data centres world wide are working towards this goal. The International Virtual Observatory Alliance (IVOA) is an organisation that debates and agrees the technical standards that are needed to make the VO possible. It was established in June 2002 with a mission to 'facilitate the international coordination and collaboration necessary for the development and deployment of the tools, systems and organizational structures necessary to enable the international utilization of astronomical archives as an integrated and interoperating virtual observatory'\footnote{http://www.ivoa.net/about/what-is-ivoa.html}. The Open Universe is another step further in this direction. 

\section{Open Universe and Education}

\begin{figure}[h]
% \vspace*{-2.0 cm}
\begin{center}
 \includegraphics[width=12.5cm, height=3.5cm]{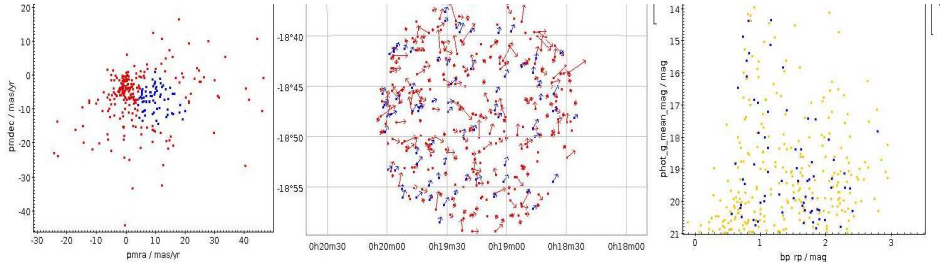} 
% \vspace*{-1.0 cm}
 \caption{Open Cluster Studies with Gaia DR2}
   \label{fig1}
\end{center}
\end{figure}
Gaia DR2 was released in April 2018\footnote{http://sci.esa.int/gaia/}.  It can be used very effectively with students using Virtual Observatory tools like Topcat and Aladin. We used it with students to make a proper motion plot to identify possible members of a star cluster. Then we made a vector plot for the members and plotted a color-magnitude diagram for the given open cluster. This could be done very easily with undergraduate students using Topcat and some simple plotting tools, including the ones offered by Topcat (Fig. \ref{fig1}). \cite[Moraux E. (2016)]{mor16} describes a variety of problems related to star clusters that can be studied using Gaia data. Interested students can be guided to work and study deeper research problems using simple initiative steps such as these. 
 We have been mentoring students on projects like these. 
 
%\begin{figure}[b]
%%%\vspace*{-2.0 cm}
%%\begin{center}
%\includegraphics[width=8.5cm]{gaia.eps} 
%%\vspace*{-1.0 cm}
%\caption{Open Cluster Studies with Gaia DR2}
%%\label{fig1}
%%\end{center}
%\end{figure}
%

\section{Conclusion}
We have shown how Open Universe and Virtual Observatory are the best options for countries which lack world-class facilities in observations. It is easy to develop computing power and train students on Data Analysism, study Statistics,  Big Data, python coding, etc.
The author thanks IAU for the travel grant that enabled participation.

\end{document}